# Thickness dependent mechanical properties of soft ferromagnetic two-dimensional CoTe$_2$

*Surbhi Slathia[†], Cencen Wei[‡], Manoj Tripathi[‡*], Raphael Tromer[Δ], Solomon Demiss Negedu[∥], Conor Boland[‡], Suman Sarkar[Φ], Douglas S. Galvao[Δ*], Alan Dalton[‡*], Chandra Sekhar Tiwary[†⁂*]*

[†]School of Nano Science and Technology, Indian Institute of Kharagpur, West Bengal, 721302, India.

[‡]Department of Physics and Astronomy, University of Sussex, Brighton BN1 9RH, United Kingdom.

[Δ]Applied Physics Department, State University of Campinas, Campinas, SP, 13083-970, Brazil.

[∥]Materials Science and Engineering, Jimma Institute of Technology, Jimma University, Jimma 378, Ethiopia

[Φ]Materials Engineering, Indian Institute of Technology Jammu, Jammu 181221, India.

[⁂]Metallurgical and Materials Engineering, Indian Institute of Technology Kharagpur, West Bengal, India.

ABSTRACT. Two dimensional (2D) layered transition-metal-based tellurides (chalcogens) are known to harness their surface atoms characteristics to enhance topographical activities for energy conversion, storage, and magnetic applications. High surface energy due to unsaturated dangling bonds and larger lateral size than the thickness (volume) makes them a potential candidate for emerging electronics. Nevertheless, the gradual stacking of each sheet alters the surface atoms' subtle features, such as lattice expansion, leading to several phenomena and rendering tunable properties. In the present work, we have monitored thickness-dependent properties of the 2D $CoTe_2$ sheets from nanoscale mechanics, tribology, surface potential distributions, interfacial interaction and magnetism using atomically resolved spectroscopy and different surface probe techniques, in conjunction with theoretical investigations: density functional theory (DFT) and molecular dynamics (MD). The variation in properties observed in theoretical investigation unleashes the crucial role of crystal planes of the $CoTe_2$. The presented results are beneficial in expanding the use of 2D telluride family in flexible electronics, piezo sensors, tribo-generator, and next-generation memory devices.

INTRODUCTION

Transition metal dichalcogenides (TMDs) have drawn significant research interest among the family of 2-Dimensional materials due to their layer-dependent properties.[1] The origin of these characteristics is due to their honeycomb structure with a unique symmetry and higher surface-to-volume ratio. One of the intriguing features of TMDs, which makes them exceptional, is their tunable bandgap, which varies as the number of layers changes from indirect at multilayer to direct at single layer, [2,3] making them potential candidates for a wide range of electronics, optoelectronics, energy storage, and catalytic applications. [4-6] Therefore, the investigation of the thickness-dependent properties of emerging TMDs have been receiving increased attention in recent years. Cao, Tianci et al. revealed that the fracture strength of graphene paper could be improved by decreasing its thickness.[7] Enhancement in film resistivity of $WS_2$ was observed with decreasing thickness by Romanov, Roman, et al..[8] The strong dependence on the layer number of $NbSe_2$ sheets was observed in their superconductivity transition temperature values.[9] In the field of nanoscale lubrication, there is clear evidence for the role of thickness over varieties of 2D materials for altering the friction force values.[10] The TMDs ($NbTe_2$, $VTe_2$, $Cr_2S_3$, etc.) are experimentally proven to exhibit magnetic behaviour when the layer thickness is decreased.[11-13] However, only a few TMDs have been explored experimentally on the nanoscale, as well-controlled layer thickness up to monolayer is preferable to study the layer-dependent properties of these materials.

Among various TMDs, ditelluride transition metals are gaining attention due to their metallic characteristics in contrast to other related structures, such as selenides and sulfides.[14] Also, tellurides exhibit unique bandgap sensitivity at lower dimensions under applied strain conditions. [15-17] 2D $CoTe_2$ is a layered material having a hexagonal structure wherein six Te atoms surround one Co atom. Several reports have experimentally proven that $CoTe_x$, in the composition range of x=1.3 up to 2 is metallic due to the decrease in the emission at the Fermi level ($E_F$) with the Te content.[18] Recent studies on $CoTe_2$ nanosheets reported that they have good electrical conductivities up to $4.0 \times 10^5$ S/m, with high breakdown current densities up to $2.1 \times 10^7$ A/cm$^2$,[19] and linear magnetoresistance up to 9T.[20] Negedu, et al. reported that $CoTe_2$ could behave as a piezoelectric material, and applying small forces (up to 1N) can result in 5V of output voltage. They further used 2D $CoTe_2$ to fabricate piezo and triboelectric nanogenerators (PTNG), which showed excellent high temperature stability and gave out an output voltage of about 10 V.[21] However, the nanoscopic size and thickness-dependent $CoTe_2$ has not yet been comprehensively investigated, and it is one of the objectives of the present study.

In the present work, the synthesis of 2D $CoTe_2$ was carried out using the liquid phase exfoliation (LPE) method through simple sonication for layer-by-layer exfoliation from the bulk $CoTe_2$. It is prepared from vacuum induction melting as it is a scalable technique and produces good quality (crystalline with a small number of impurities) nanomaterials at a reasonable cost.[22] X-ray diffraction (XRD) technique and microscopy images confirmed the formation of 2D layered $CoTe_2$ nanoflakes. The produced 2D $CoTe_2$ was extensively investigated through atomically resolved transmission electron microscopy to determine the lattice spacing as a function of the number of layers. The samples were further investigated through nano mechanics (mechanical mapping), tribo-mechanics (lateral force microscopy), surface potential distribution (KPFM) and nanoscale magnetism. We also carried out fully atomistic computer (DFT and *ab initio* molecular dynamics) simulations to better interpret the experimental data and associated physicochemical mechanisms. The obtained results add important information on the emerging class of tellurides and their use in applications for next-generation electronic devices.

EXPERIMENTAL SECTION

**Synthesis.** Bulk $CoTe_2$ was prepared with vacuum induction melting method by using cobalt and tellurium (99.99% purity) in a stoichiometric ratio of 17.5 wt% of Cobalt and 82.5 wt% of tellurium for the composite formation. The alloy was made by melting the elements in a quartz tube at a temperature of 1050 °C in a melting chamber under an argon atmosphere. The melting chamber was kept at a high vacuum condition of $10^{-5}$ mbarr for obtaining pure alloy and preventing oxidation. Then, the prepared composite was furnace cooled and then converted to powder using mortar and pestle. From these powdered bulk samples, 2D $CoTe_2$ samples were prepared by taking 50 mg of the bulk sample in 150 ml of isopropyl alcohol and ultrasonicated in a probe sonicator (f =30 kHz) for 3 hours at room temperature to obtain exfoliated $CoTe_2$ sheets.

**Instrumentation.** The diffraction patterns and crystalline phase information of the sample were taken from Bruker, D8 Advance XRD with Cu-kα radiation having 1.5406 Å wavelength (λ) with 40 kV voltage and 40 mA current as operating conditions. WiTec UHTS Raman spectrometer 300 VIS, Germany having an excitation wavelength of 532 nm was used for Raman spectrum analysis at Room temperature. To study the surface composition and oxidation states of exfoliated $CoTe_2$, the PHI 5000 Versa probe-III scanning XPS microprobe was used (SI **Figure S6**). To study topographical, compositional, and crystalline properties FEI, Themis 60-300, and FEI- Ceta 4k×4k camera HRTEM were used.

Atomic Force Microscopy (AFM) was used for nanoscale imaging, potential mapping, lateral force microscopy and magnetic force microscopy. All these characterizations were carried out using the Bruker Dimension Icon instrument at room temperature with 35% relative humidity. The AFM was positioned in the insulated box over an anti-vibrant stage to mitigate the environmental noise and building vibrations. For each mode specific cantilevers were used for measurements as follows:

- AFM was carried out with quantitative nanomechanical mapping (QNM) mode, "an areal collection of force-displacement spectroscopy" for measuring the variation in mechanical properties such as modulus, adhesion and indentation along the thickness changing from monolayer to bulk. For the measurements, probe stiffness, driving frequency, and tip radii were taken at values of 5±0.5 N/m, 150 kHz, and 10±2 nm, respectively. For calibration of the cantilever, -Sader's method,[23] as well as thermal tuning, were used. The uniqueness of PF-AFM in tapping mode is that the maximum normal force applied at each point on the sample surface can be controlled using a feedback system that keeps the maximum force constant while scanning. Silicon nitride probes and Sn doped Si probe (Model: RTESPA-150, Bruker) with stiffness and drive frequency of 5±0.5 N/m and 150 kHz approximately, and tip radii of 10±2 nm were used for the measurement.

- Lateral force microscopy (LFM): To measure the lateral force by sliding the tip apex in trace and retrace direction a silicon cantilever of stiffness 0.4 N/m (Model: CSG10, NT-MDT) was used.

- Kelvin probe force microscopy (KPFM): Peak force-Kelvin probe microscopy (Model: PFQNE-AL) to measure Contact potential difference (CPD, volts)

- Magnetic force microscopy (MFM): Bruker MESP probes (cobalt-chromium coating on tip apex) has been used for the investigation. The influence of the localized magnetic field has been measured through phase image in interleave mode.

**Preparation of sample for AFM analysis.** The 2D $CoTe_2$ powdered sample was mixed into 10 ml of IPA and ultrasonicated for up to 30 minutes. The dispersion was spin-coated over the Si wafer at 1000 rpm for 1 minute. The prepared sample was further used for the doing AFM analyses.

**DFT Simulations.** In order to obtain further insights and to better interpret the experimental results, we carried out ab initio calculations at the density functional theory (DFT) level. We use the cut-off energy of 250 Ry and k-points mesh of 2x2x2 within the local density approximation functional (LDA) and Ceperley and Alder (CA) parametrization [24] and single zeta SZ basis. We assume that when forces in each atom are less than 0.05 eV/Å, the convergence criteria to the optimization process were achieved. All calculations were performed with SIESTA software.[25]

We considered, in our theoretical analyses, structural models to mimic the experimental conditions, i.e., the $CoTe_2$ layers were deposited on silicon oxide (silica, $SiO_2$). We deposited at the substrate slab afragments of 2D $CoTe_2$ corresponding to from one up to four layers at three different crystallographic orientations: [001], [010], and [001]. The heterolayer structures are then fully geometrically optimized.

Once we obtained these optimized structures, we carried out two types of analyses: i) the energy layer adhesion, and; ii) a TIP indentation on the 2D $CoTe_2$ layers. The TIP was modelled using a triangular $Si_3N_4$ shaped/pyramid, which is the same tip material used in the AFM measurements. The adhesion of layers was calculated considering the work necessary to extract each layer from different sizes, using the expression (1):[26]

$$W_{ad} = \frac{E_{SiO_2} + E_{2DCoTe_2} - E_{\frac{2DCoTe_2}{SiO_2}}}{A} \quad (1)$$

where $E_{SiO2}+E_{2DCoTe2}$ are total energy values for the isolated $SiO_2$ and 2D $CoTe_2$, respectively, $E_{2DCoTe2/SiO2}$ is the total energy value for the 2D heterostructure $CoTe_2/SiO_2$, and A is the contact area between them.

From equation (1), we calculated the work of adhesion ($W_{ad}$) for different separation values of sample and substrate. The $W_{ad}$ values were estimated from the universal binding energy relation curves (UBER), [26] which is a fitting process to calculate the work of adhesion as a function of 2D $CoTe_2$ number layers and their separation distance from the silica substrate.

For the analysis of $Si_3N_4$ TIP penetration on the 2D $CoTe_2/SiO_2$ heterostructure, we performed ab initio molecular dynamics (AIMD) simulations with an NVT ensemble and integration time of dt=1.0 fs. We simulated the TIP processes assuming indentation step movements of 0.5 Å, in which, for each step, an AIMD run of 0.5 ps is performed. When the Si3N4 TIP penetrates the $CoTe_2$, it produces a compressive stress (negative value) along the Z direction. It relates to the force through the expression:

$$S_Z = -\frac{F}{A} \quad (2)$$

where $S_Z$ is the stress component along the Z direction, F is the force applied to the system and A is the contact area at which the force is applied.

RESULTS AND DISCUSSION

The schematic representation of the formation of layered 2D $CoTe_2$ sheets from its bulk counterpart described in **Figure 1a**. The diffraction pattern of the obtained 2D $CoTe_2$ is shown in **Figure 1b**, indexed to non-centrosymmetric space group Pnn2 (No. 34) having lattice parameters of a = 5.329Å, b = 6.322Å, and c = 3.908Å. The peaks were observed at $28.3^0$, $31.7^0$, $32.9^0$, $47.2^0$, $54.31^0$, and $58.2^0$, which are characteristic peaks of orthorhombic crystalline structures. High-resolution transmission electron microscopy (HRTEM) images of the polycrystalline telluride flakes of different thicknesses are presented in **Figure 1c**. The atomic resolution identifies different crystallographic plane orientations (**Figure 1d**) with different atomic arrangements. A zoom-in region of a particular plane in **Figure 1e** confirms the presence of two types of atoms: Co and Te, which are represented by white and black balls, respectively. The Co atomic number is smaller than Te's, thus, it shows a weak contrast compared to Te. The positions of Co and Te are shown using blue and yellow colors, respectively (**Figure 1e**), which reveals the $CoTe_2$ hexagonal structure. The occurring formation of structures with a different number of layers (consequently, having different contrasts) can be clearly seen in **Figure 1f**. The insets show the FFT and inverse FFT analysis of the plane orientation observed in these different layers. Further, the $d_{spacing}$ was calculated for the different layers of the $CoTe_2$ nanoflake, indicating a gradual increase with thin layers having $d_{spacing}$ value of approximately 0.15 nm and thick layers having 0.22 nm. This plays a crucial role in influencing the mechanical properties, surface chemistry, surface potential, and magnetic properties discussed in the subsequent sections.

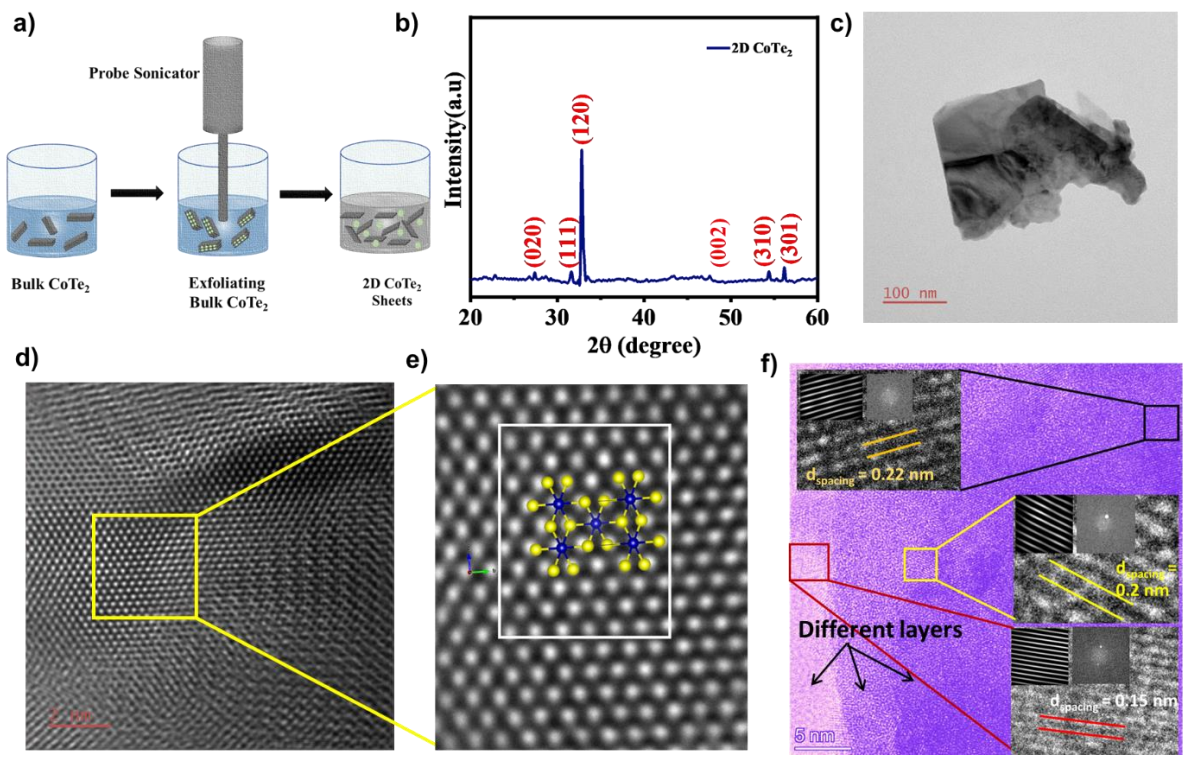

**Figure 1.** Structural and morphological characterization of 2D $CoTe_2$. (a) Schematic representation of synthesis procedures of 2D $CoTe_2$, (b) X-ray diffraction pattern of 2D $CoTe_2$, (c)

HRTEM image of a CoTe$_2$ flake, (d) Different crystallographic orientations observed in the HRTEM image, (e) Magnified image distinguishing Co and Te atoms and (f) TEM image showing different layers of 2D CoTe$_2$ nanoflake (magnified view of different crystal plane orientations, the insets show FFT and inverse FFT analyses).

The morphology of the 2D CoTe$_2$ flakes of different thicknesses and their corresponding mechanical response against calibrated tip apex was investigated through atomic force microscopy (AFM); see experimental section for calibration of cantilevers. The nanoscale mapping through peak tapping probe microscopy (PF-AFM) is useful to obtain the mechanical properties of the telluride sheets under elastic conditions (i.e., without damage) of varying thickness from 1 up to 9 nm, **Figure 2a and b**. Thus, the concurrent maps of modulus, adhesion and indentation were obtained in a single acquisition for a valid comparison, assuming unaffected tip radii. The Derjaguin-Muller-Toporov (DMT) [27] contact mechanical model was used, as it is applicable for stiff samples having low surface energies and includes van der Waals forces outside the contact region. As per this model, the tip-sample surface interaction force under the influence of adhesion force can be given by the relation *(3)*:

$$F = \frac{4}{3} E^* \sqrt{R(d - d_o)^3} + F_{adhesion} \qquad (3)$$

Where R is the tip radius, d$_o$ is the surface rest position, (d-d$_o$) is the sample deformation. E$^*$ is said to be the effective elastic modulus of tip and sample, and F$_{adhesion}$ is the adhesion force during contact. The contact mechanics in the DMT conditions is useful to obtain the mechanical properties over various thickness of 2D CoTe$_2$ shown in **Figure 2b**. The retraction force-displacement curves for 2D CoTe$_2$ nanoflakes from monolayer to bulk are illustrated in **Figure 2c**. From these retraction force calculations, we observed a gradual reduction in the adhesion force "pull-out" and an increase in indentation depth from layer thickness 1 nm to 9 nm, a common softness behavior as a function of the thickness of van der Waals solids.

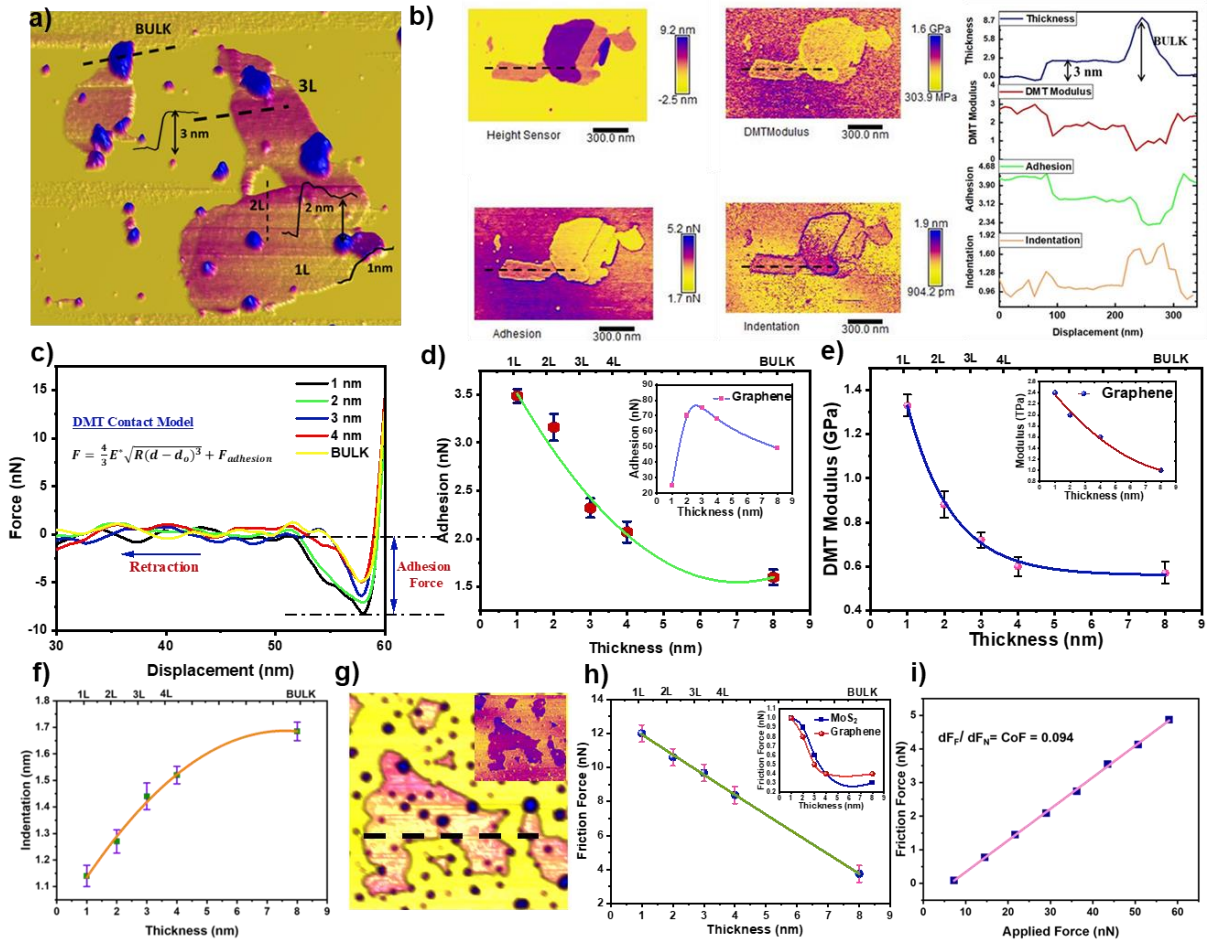

**Figure 2.** The concurrent surface chemistry (adhesion force) and mechanical response (DMT modulus, indentation map) from $CoTe_2$ sheets. (a, b) AFM topography of 2D $CoTe_2$ layers distributed from monolayer to bulk at different locations. The dashed line shows the thickness of an individual sheet and its mechanical response towards the AFM tip apex sheet quantified through the line profile. (c) The force-displacement retraction curves from a few layers to bulk show an adhesion force (pull-out) from different thicknesses. The inset presents the contact-mechanical model used to measure the mechanical properties under the influence of the adhesion forces. (d, e, f) Variation in average values of adhesion, modulus, and indentation as a function of the thickness.

The insets show a comparison with graphene. [28,29] (g, h) Frictional map and its characteristics at different thicknesses compared to other 2D materials (inset [34]), and (i) Calculation of the friction coefficient from the applied force Vs. friction force graph of a fixed thickness.

**Figure 2 d, e and f** summarizes the trend of average values of adhesion force (nN), modulus (GPa) and indentation depth (nm) values from several telluride sheets showing decreasing adhesion force, modulus, and increasing indentation depth at fixed normal force value. The adhesion force values depend on two major factors: the interfacial contact area between the tip apex-topmost layer of the 2D material under deformation and the interfacial interaction through static electric charges. The indentation depth represents the separation between jump-to-contact and maximum force point, previously referred to as deformation-distance in expression (3). The increment of indentation depth (nm) indicates the higher contact area with increasing thickness results in higher adhesion force as observed in graphene (see inset **Figure 2e**). Nevertheless, the lowering of adhesion force in the telluride sheets confirms the predominance of the static interfacial charges for determining the interfacial interaction (i.e., the metallic tip experiences a repulsion force from the thicker telluride sheets). Further, the increment of the indentation depth also explains the softness (1L: 1.4±0.5 GPa, 2L: 0.8±0.5 GPa and bulk: 0.6±0.5 GPa) with increasing thickness, similar to the graphene sheets (see the inset of **Figure 2f**) of varying from Young's modulus from 2.4 TPa to 1 TPa from monolayer to bulk.[29,30] The additional factor responsible for mechanical characteristics with thickness is increasing the lattice constant, which leads to weak binding between metal and chalcogenides. During indentation, the interlayer stacking error, i.e., interlayer sliding in multilayer 2D materials, can also be a factor for the decrease in modulus values with increasing thickness. [31] Thus, 2D $CoTe_2$ sheets share a similar mechanical response with graphene layers with distinct interfacial interaction that will be discussed through the surface potential map by kelvin probe force microscopy (KPFM) in the subsequent section.

One of the advantages of having low adhesion forces is the potential use of the 2D $CoTe_2$ sheets as a solid-state lubricant and tribo-generator.[21] The friction map generated from shear forces between the sliding tip and the telluride sheets shows lower resistance to sliding/delamination, **Figure 2g** (inset concurrent friction map). The average values of friction force (nN) decrease with thickness as shown in **Figure 2h.** This tribological trend is in good agreement with the pioneer works of Lee et al.[10] and Filleter et al.,[32] addressing the frictional behavior of 2D materials for different thicknesses. This trend is observed (although with different frictional values) for the majority of 2D materials, including graphene and $MoS_2$;[33-35] see inset **Figure 2h**. It has been proposed that thin layers of a 2D material with weak interlayer and substrate interactions are susceptible to deform out-off plane ("a puckering effect") under shear forces, as depicted in schematic **Figure S2** in the supplementary information (SI). When the AFM tip slides over the surface of the sample with a finite normal force, a puckered structure (i.e., temporary wrinkle-like topology) is formed around the tip, which increases the contact area and thus induces higher friction. However, as the number of layers is increased, the bending rigidity increases along the vertical direction, which results in suppressing the puckering, and consequently lowering the friction force. A generalized form of nanoscale frictional feature can be described through the coefficient of friction (CoF) values through a first derivative of load-dependent friction. The estimated value of 0.094 bring tellurides close to the category of graphene and $MoS_2$ as good solid-state lubricants.

Density functional theory (DFT) and *ab initio* molecular dynamics simulations were carried out using structural models to mimic the experimental conditions of the contact interaction of the 2D $CoTe_2$ layers with a $SiO_2$ (silica) substrate. We used a $Si_3N_4$ shaped/pyramid tip (the same tip material used in the AFM measurements) to simulate the indentation processes. We analyzed the adhesion energy variations and the structural distortions of the 2D layers during the simulated indentations. See experimental section for details.

The adhesion energy ($J/m^2$) is measured as a function of the separation distance (Å) between the telluride sheet/s (1L and 2L) against the $SiO_2$ surface (see **Figure 3a and b** for the telluride crystal plane [010]). The optimized structures showed that the charge in the $SiO_2$ substrate remains practically the same, and no significant charge transfer has been observed from $CoTe_2$ to the $SiO_2$ substrate. A slight decrease in the adhesion energy value from 1L (1.7 $J/m^2$) to 2L (1.66 $J/m^2$) corroborates the trend observed from force-distance spectroscopy measurements. The in-depth analysis indicates that negative static charges on the interface of $CoTe_2/SiO_2$ (26 me) decrease as a function of the number of layers and plays a crucial role in determining the adhesion energy value. It is important to remark that this, trend of adhesion energies has been observed for other crystal planes [100] and [001], see SI (**Figure S5**), which suggests that the above-mentioned trend is crystal-oriented dependent and consistent with the different static charges values 126 and 35 meV along [100] and [001] plane respectively. Also, we considered our structures in vacuum and defectless, while the experiments were in an aqueous environment (35% of relative humidity) and contained structural defects, as evidenced by the microscopy results. Broadly, the adhesion energy of $CoTe_2$-$SiO_2$ depends on the crystal plane interfacing the silica substrate, and lies in the range of 0.97 to 1.70 $J/m^2$, which is similar to other 2D materials, such as 1L graphene- $SiO_2$ (0.72 $J/m^2$),[36] $MoS_2$- $SiO_2$ (0.17 $J/m^2$).[37] Graphene has shown a parabolic trend in the adhesion energy from 1L-3L (2L: 1.4 $J/m^2$ and 3L: 1.3 $J/m^2$), indicating saturation of one type of charge at the bilayer.[38]

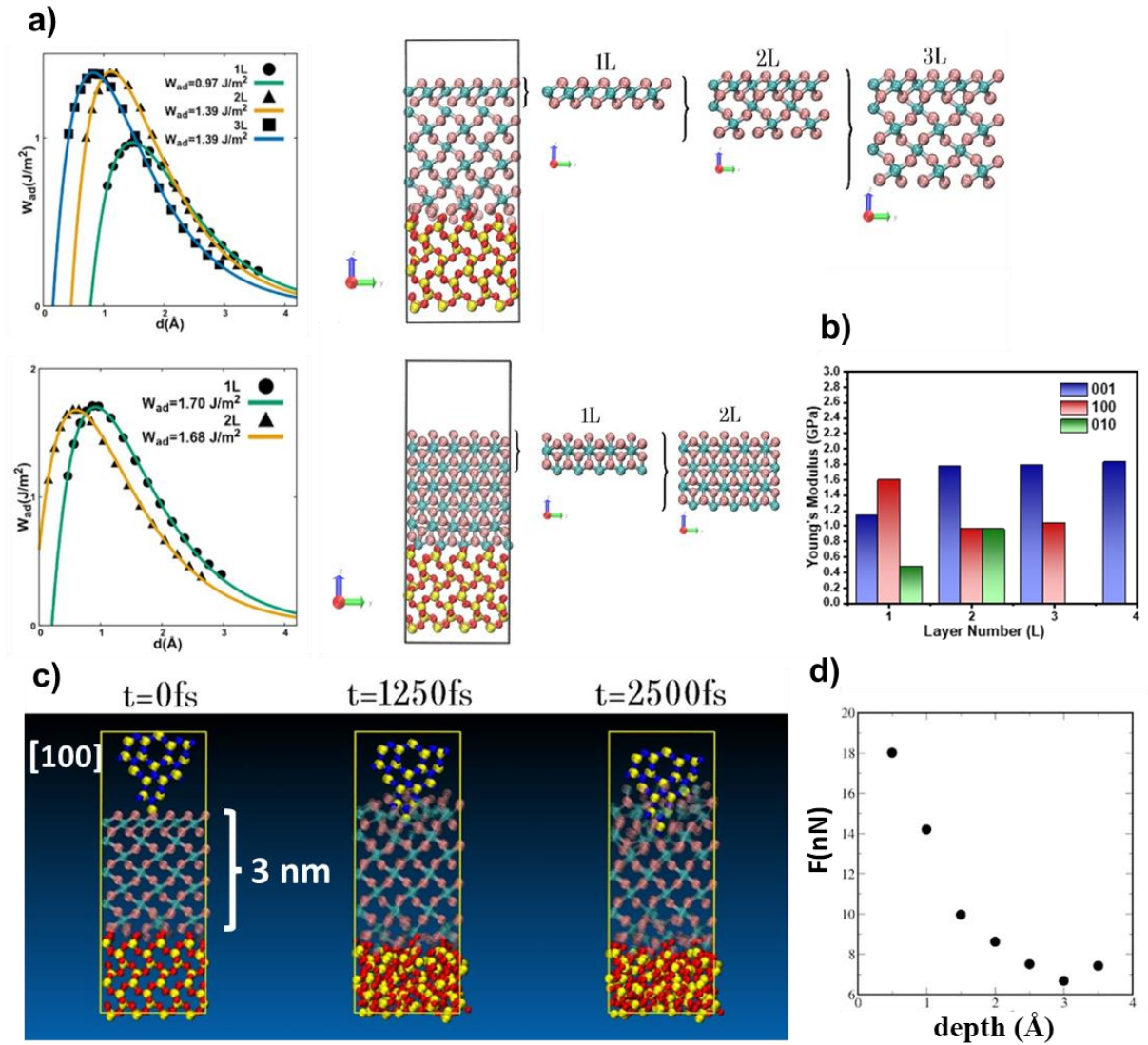

**Figure 3.** Schematic of structural models (lateral views) of the 2D CoTe$_2$/SiO$_2$ interfaces (we considered from a single layer (1L) up to four layers (4L). (a) The adhesion energy values, as a function of the separation distance for the [100] and [010] cases. (b) Young's modulus values as a function of the number of layers for the crystallographic directions considered here. (c) Indentation process of three layers (3L) at different time frames (0 fs, 1750 fs and 3500 fs) from the event of "jump-to-contact" to a finite indentation depth. (d) Indentation force (nN) as a function of the depth value (Å).

We also estimated Young's modulus ($Y_M$) values from the DFT simulations. The procedure is to introduce small strain values (~1%) along the X and Y directions separately and then obtain the $Y_M$ values from the slope of the stress versus strain curves. The $Y_M$ values for different crystallographic directions as a function of the number of layers are presented in **Figure 3b**. As can be seen from this Figure, the different crystallographic directions do not show the same trends. The obtained numbers are consistent with the corresponding experimental ones. The simulated indentation procedures were performed through ab initio molecular dynamics (AIMD) simulations. We considered a rigid indenter ($Si_3N_4$) that is moved (perpendicularly) towards the $CoTe_2/SiO_2$ structure up to a penetration depth of 0.5 Å, as schematically shown in **Figure 3c** for the case of $CoTe_2$ [100] plane. The corresponding AIMD snapshots at different time frames (fs) show a gradual penetration of the indenter and accumulation/dissipation of induced stress through each layer to the silica substrate, which is different for each crystallographic direction. The corresponding results for the [010] and [001] directions are presented in the SI based on the atomic charge distribution. [100] undergoes a minor deformation (up to the depth of 3.5 Å) due to an efficient stress transfer to the silica substrate (silica amortization). **Figure 3d** presents the force as a function of the depth values. We can see from this Figure that the general trend is that the force decreases as the depth value increases. The same behavior was observed for the other crystallographic directions (results presented in the SI). Although the values from the simulations are larger than those experimentally obtained (expected as the simulations considered pristine defect less structures), the qualitative behavior is the same experimentally observed and further validates the thickness softness effect.

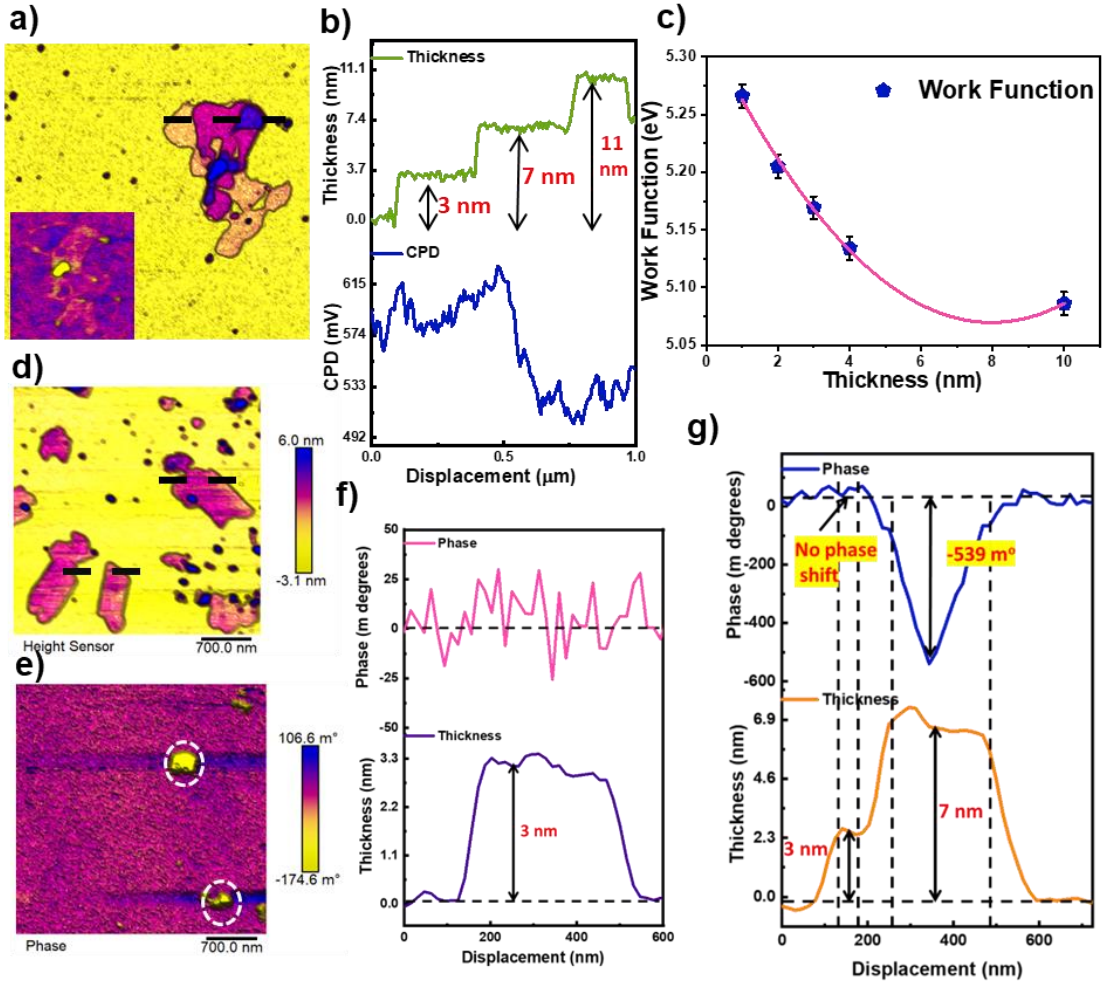

**Figure 4.** Work function and MFM measurements of CoTe$_2$ sheets. a) Topographical image of CoTe$_2$ nanoflakes, (b) Variation of potential in comparison to thickness along the displacement direction, (c) Work function values as a function of the number of layers, (d) AFM topography of CoTe$_2$ sheets obtained using a magnetic sensor, (e) Phase image susceptible to the stray magnetic field at the lift height of 20 nm of the same region illustrates selected regions (marked by dashed circles) in the CoTe$_2$ sheet. (f) Line profile over CoTe$_2$ sheet of 3nm height showing no influence from the stray magnetic field in the phase signal, and (g) Line profile from the 7 nm thick sheet and its concurrent phase profile revealing significant phase shift from the region of CoTe$_2$.

As discussed above, the thickness dependence of 2D CoTe$_2$ sheets influences its mechanical properties and interfacial adhesion energy. Its critical effect also applies to surface potentials and magnetic fields, sensed by separate functionalized probes of metallic and magnetized coating (cobalt-chromium) coatings, respectively. The contact potential difference (CPD, volts) observed during peak force (PF)-KPFM measurements shows topography corresponding to surface potential distribution of different thicknesses, **Figure 4a and b**. The average values of work function (eV) were calculated from surface potential values using an expression (SI Equation 4) showing a decreasing trend with thickness, **Figure 4c**. The decreasing values of work function with increasing thickness indicate higher electron concentration at the thick layer. This phenomenon can be explained by electrostatic interlayer screening, [39] i.e., charge redistribution among each layer induced from the substrate as observed in graphene by Kim and co-workers. [40] Therefore, increasing the negative charges with the addition of a telluride layer repels the similar negative charges on the sliding probe responsible for the lower adhesion with higher thickness nevertheless, interlayer interaction (in thick sheet) against tip apex cannot be ignored.[41] One can stimulate the surface with biasing and modulate the static charge interaction, consequently, adhesion force from 2D material for desired application.[42]

The magnetostatic interaction between the stray magnetic field from the CoTe$_2$ and magnetic sensor was investigated as the function of thickness. The magnetized probe (see experimental section) is slid at constant lift height to detect the force gradient of the stray field. The vertical force gradient from different sheet thicknesses shifts the magnetic sensor resonance frequency ($f_o$) proportionally. The frequency shift ($\Delta f$) is detected through the phase of oscillations. In the present set-up, the cantilever phase signal upper shift indicates the magnetic sensor repulsion from the stray magnetic field and vice-versa for the downshift (see scale bar in **Figure 4e**). The phase image reveals that a limited region of the CoTe$_2$ sheet is susceptible to the magnetic field, which is quantified through the phase signal profile. It is observed that thicker CoTe$_2$ (around 7 nm) showed the signature of the magnetic attraction, which was not observed in the thinner sheets of 3 nm (See **Figure 4f and g**). It is worth mentioning that MFM technique is well suited for highly magnetized samples, nevertheless ineffective towards weak stray fields generated from subtle magnetic regions.[43] DFT simulations were also carried out to investigate sub-nanoscale magnetic properties of 2D CoTe$_2$ through the calculation of collinear spin polarization (see SI, **Figure S5** for details). It was observed that magnetic moment ($\mu$) increased with thickness for the crystal direction [100], as follows: ($\mu_{1L}$= 0.43 $\mu_B$ , $\mu_{2L}$= 0.56 $\mu_B$, and $\mu_{3L}$= 0.58 $\mu_B$ ). For the other crystal planes ([010], and [001]), different trends were observed ($\mu_{1L}$= 0.62 $\mu_B$ , and $\mu_{2L}$= 0.57 $\mu_B$), ($\mu_{1L}$= 0.77 $\mu_B$ , $\mu_{2L}$= 0.62 $\mu_B$, $\mu_{3L}$= 0.57 $\mu_B$), respectively, where, $\mu_B$ is the Bohr magneton constant. The DFT results also indicate that the magnetization of CoTe$_2$ is direction dependent, it increases for [100], and decreases to [010] and [001].

CONCLUSION

In summary, the present work demonstrated the thickness-dependent properties of CoTe$_2$, which is an emerging material in 2D-materials families. We have observed structural changes as a function of the number of layers and at the different crystallographic orientations, which are responsible for mechanical response, tribology, surface charge distribution, and nanoscale magnetism. The obtained results for nanostructured CoTe$_2$ were contrasted with literature ones for graphene and MoS$_2$, in order to better evaluate the potential of its applications. We have demonstrated that the number of layers can be effectively used to tune and control some electronic and/or magnetic features. The experimental and DFT obtained results add important information for tellurides, that can be helpful for designing next-generation Tellurides-based devices.

**Notes**

The authors declare no competing financial/commercial conflicts of interest.

ACKNOWLEDGEMENTS

M.T. and A.B.D. would like to thank Sussex strategic development funds to carry out research at nanoscale.

ABBREVATIONS

DFT, density functional theory; TMDs, transition metal dichalcogenides; XRD, X-ray diffraction; XPS, X-ray photoelectron spectroscopy; HRTEM, high-resolution transmission electron microscopy; AFM, atomic force microscopy; KPFM, kelvin probe force microscopy; LFM, lateral force microscopy; MFM, magnetic force microscopy; AIMD, ab initio molecular dynamics.